\documentclass[journal]{IEEEtran}

\usepackage{amsmath}
\usepackage{amssymb}
\usepackage{graphicx}
\usepackage{booktabs}
\usepackage{multirow}
\usepackage{makecell}
\usepackage[table]{xcolor}   
\usepackage{url}
\usepackage[hidelinks]{hyperref}
\usepackage[utf8]{inputenc}
\usepackage{balance}   

\definecolor{lightred}{RGB}{255,230,230}

\newif\ifshowtodo
\showtodotrue

\begin{document}

\title{TinyCNNDeep: Lightweight Attention-Based CNN for EEG Classification of Eye States and Sleep Deprivation}

\author{Thien~Nhan~Vo,~Yen~Nhi~Tran~Thi,~Ngan~Nguyen~Xuan~Phuong~and~Xuan-The~Tran%
\thanks{T. N. Vo is with the HAI-Smartlink Research Lab, Anchi STE Company, Haiphong, Vietnam, and with the HUTECH Institute of Engineering, HUTECH University, Ho Chi Minh City, Vietnam.}%
\thanks{Y. N. Tran Thi is with the HUTECH Institute of Engineering, HUTECH University, Ho Chi Minh City, Vietnam.}%
\thanks{N.N Xuan Phuong is with the  Institute of Intelligent \& Interactive Technologies - University of Economics Ho Chi Minh City (UEH), Ho Chi Minh, City, Viet Nam }%
\thanks{X.-T. Tran is with the School of Mechanical Engineering, Vietnam Maritime University, Haiphong, Vietnam.}%
\thanks{Corresponding author: Xuan-The Tran (thetx.vck@vimaru.edu.vn)}%
\thanks{This work has been submitted for possible publication. Copyright may be transferred without notice, after which this version may no longer be accessible.}}

\markboth{Preprint, 2026}%
{Vo \MakeLowercase{\textit{et al.}}: TinyCNNDeep for EEG-Based Eye-State Classification Under Sleep Deprivation}

\maketitle

\begin{abstract}
Sleep deprivation impairs vigilance and cognitive function, yet jointly identifying the sleep condition (normal vs. deprived) and the eye state (open vs. closed) from electroencephalography (EEG) remains underexplored. We address this four-class problem with TinyCNNDeep, a lightweight convolutional neural network that combines residual learning with a Squeeze-and-Excitation (SE) attention module. We convert short multi-channel EEG segments from five physiologically relevant channels (Fp1, Fp2, O1, Oz, O2) into $224\times224$ grayscale images through per-channel Z-score normalization, min-max scaling, and center padding, enabling 2D convolutions to jointly model inter-channel and temporal structure. On a 35-subject dataset recorded under normal-sleep and sleep-deprivation sessions, TinyCNNDeep attains a subject-wise mean accuracy of 83.69\%, outperforming the strongest baseline (Random Forest with combined time-frequency features, 47.66\%) by 36.03 percentage points, while three established EEG architectures (EEGNet, ShallowConvNet, DeepConvNet) operate near chance. Per-subject analysis quantifies inter-subject variability, and confusion-matrix inspection shows that residual misclassifications concentrate between eyes-closed states across sleep conditions. These results indicate that an image-based EEG representation paired with residual feature extraction and channel attention provides an accurate and computationally efficient framework for multi-class sleep-related EEG classification under a minimal electrode configuration.
\end{abstract}


\IEEEpeerreviewmaketitle

\section{Introduction}
\label{sec:introduction}
\IEEEPARstart{S}{leep} is a fundamental biological process essential for cognitive function, memory consolidation, emotional regulation, and overall physiological homeostasis~\cite{walker2017sleep, diekelmann2010memory}. Sleep deprivation (SD), defined as obtaining insufficient sleep relative to individual needs, has become increasingly prevalent in modern society due to occupational demands, lifestyle factors, and medical conditions~\cite{chattu2019insufficient}. Chronic and acute sleep deprivation have been consistently associated with impaired attention, reduced reaction time, diminished working memory capacity, and increased risk of accidents in safety-critical domains such as transportation and healthcare~\cite{lim2010meta, killgore2010effects}.
Electroencephalography (EEG) has long served as the gold standard for non-invasive monitoring of brain electrical activity and has been extensively employed in sleep research~\cite{berry2017aasm}. EEG signals capture the spatiotemporal dynamics of cortical oscillations, providing rich information about vigilance states, arousal levels, and cognitive engagement~\cite{borbely2016two}. Of particular interest is the distinction between eyes-open (EO) and eyes-closed (EC) resting states, which exhibit well-characterized neurophysiological differences. The EC condition is typically dominated by alpha-band (8--13~Hz) oscillations, particularly over occipital regions, while the EO condition is characterized by alpha suppression and increased beta-band activity~\cite{barry2007eeg}. Sleep deprivation modulates these patterns by inducing complex changes in spectral power, including alterations in alpha bandwidth and aperiodic activity, which are closely linked to declining vigilance~\cite{lorenzo1995tsd, hoedlmoser2024alpha, cassim2024aperiodic}.
Traditional approaches to EEG-based sleep state classification have relied on manually engineered features extracted from time-domain, frequency-domain, or time-frequency representations, subsequently fed into conventional machine learning classifiers such as Support Vector Machines (SVM), Random Forests, and k-Nearest Neighbors (KNN)~\cite{boostani2017comparative, aboalayon2016sleep}. While these methods have demonstrated moderate success, they are inherently limited by the quality and generalizability of hand-crafted features, which may fail to capture the complex, non-linear patterns embedded in EEG signals~\cite{craik2019deep}.
The advent of deep learning has significantly advanced EEG analysis by enabling automatic feature extraction directly from raw or minimally preprocessed signals. Several architectures have been specifically designed for EEG classification. EEGNet~\cite{lawhern2018eegnet} introduced a compact CNN with depthwise and separable convolutions tailored for brain-computer interface (BCI) applications. ShallowConvNet and DeepConvNet~\cite{schirrmeister2017deep} adapted filter bank common spatial patterns and deep architectures, respectively, for decoding motor imagery and other EEG paradigms. While these models have become standard benchmarks, recent studies have shown that their performance is highly variable across datasets and evaluation protocols, with no single architecture consistently dominating~\cite{hernandez2024eeg}.
More recently, attention mechanisms have emerged as powerful components for enhancing EEG classification. The Squeeze-and-Excitation (SE) mechanism~\cite{hu2018squeeze}, originally proposed for image recognition, adaptively recalibrates channel-wise feature responses and has been successfully integrated into EEG-based BCIs to emphasize task-relevant neural information while suppressing noise~\cite{li2024se_eeg, zhang2025se_bci}. Concurrently, there has been growing interest in converting EEG signals into two-dimensional image representations and leveraging the feature extraction capabilities of CNN architectures designed for computer vision tasks~\cite{bashivan2016learning, wang2024eeg_image}. This approach offers the advantage of exploiting well-established image processing techniques while preserving the spatial and temporal structure of multi-channel EEG data.
Despite these advances, several critical gaps remain in the literature. First, most existing studies focus on binary classification (e.g., sleep vs. wake, or EO vs. EC in isolation) without simultaneously discriminating between sleep conditions (normal vs. deprived) and eye states~\cite{alsolai2024sleep}. Second, many deep learning approaches require substantial computational resources, limiting their applicability in resource-constrained or real-time scenarios~\cite{phan2024sleeplight}. Third, the challenge of high inter-subject variability in EEG signals continues to limit robust cross-subject classification performance~\cite{wu2024transfer, tran2026variability}.
To address these limitations, this study proposes TinyCNNDeep, a lightweight CNN architecture that integrates residual learning with SE attention for the four-class classification of EEG signals: eyes-open and eyes-closed under both normal sleep and sleep deprivation conditions. The key contributions of this work are as follows:
\begin{enumerate}
    \item We propose an EEG-to-image conversion pipeline that transforms multi-channel EEG signals into compact grayscale images suitable for CNN-based classification, using a reduced set of five physiologically relevant channels (Fp1, Fp2, O1, Oz, O2).
    \item We design TinyCNNDeep, a lightweight architecture combining a stem block, three residual blocks with progressive channel expansion (32$\rightarrow$64$\rightarrow$128$\rightarrow$256), an SE attention module, and a compact classification head, achieving high accuracy with minimal computational overhead.
    \item We conduct comprehensive benchmarking against three established deep learning models (EEGNet, ShallowConvNet, DeepConvNet) and five classical machine learning algorithms across three feature domains (time, frequency, and combined), showing that the proposed approach attains substantially higher accuracy.
    \item We perform subject-wise evaluation on 35 subjects to provide a realistic assessment of classification performance, characterizing both the model's strengths and the impact of inter-subject variability.
\end{enumerate}
The remainder of this paper is organized as follows. Section~\ref{sec:methodology} describes the dataset, preprocessing pipeline, EEG-to-image conversion, and the proposed TinyCNNDeep architecture. Section~\ref{sec:results} presents the experimental results and benchmark comparisons. Section~\ref{sec:discussion} provides a detailed discussion of the findings in the context of related work. Section~\ref{sec:conclusion} concludes the paper with a summary and future directions.

\section{Related Work}
\label{sec:relatedwork}
We organize prior work into four threads that the proposed method draws upon: classical feature-based EEG classification, deep learning architectures for EEG, attention mechanisms and image-based EEG representations, and sleep-deprivation detection.

\subsection{Classical Feature-Based EEG Classification}
Early work on EEG-based sleep and vigilance analysis relied on hand-crafted features extracted from time-domain, frequency-domain, or time-frequency representations, which were then classified by conventional algorithms such as SVM, Random Forest, and KNN~\cite{boostani2017comparative, aboalayon2016sleep}. These pipelines are interpretable and data-efficient, but their performance is bounded by the discriminative power of the chosen features, which often fail to capture the non-linear, subject-dependent structure of EEG signals~\cite{craik2019deep}. We adopt five such classifiers across three feature domains as baselines in Section~\ref{subsec:baselines}.

\subsection{Deep Learning Architectures for EEG}
To remove the dependence on manual feature engineering, compact convolutional networks have been designed specifically for EEG. EEGNet~\cite{lawhern2018eegnet} uses depthwise and separable convolutions to model spatial and temporal structure with few parameters, while ShallowConvNet and DeepConvNet~\cite{schirrmeister2017deep} emulate filter-bank common spatial patterns and deeper hierarchical decoding, respectively. Although these models are standard benchmarks, recent benchmarking studies report that their accuracy is highly sensitive to dataset, channel count, and training-data volume, with no single architecture dominating across paradigms~\cite{hernandez2024eeg}. Beyond convolutional designs, alternative deep architectures such as state-space models have been explored for EEG classification tasks including dementia detection~\cite{tran2024eegssm}. This motivates evaluating whether established convolutional architectures transfer to the four-class, five-channel setting studied here.

\subsection{Attention Mechanisms and Image-Based EEG Representations}
The Squeeze-and-Excitation (SE) mechanism~\cite{hu2018squeeze} recalibrates channel-wise feature responses and has been integrated into EEG networks to emphasize task-relevant components and suppress noise~\cite{li2024se_eeg, zhang2025se_bci}. Attention has also been combined with neural memory for long-horizon EEG tasks such as seizure forecasting~\cite{pham2026eegtitans}. In parallel, a growing line of work converts EEG into two-dimensional representations so that CNN architectures from computer vision can be applied~\cite{bashivan2016learning, wang2024eeg_image}. Prior image-based methods typically use full-scalp topographic maps or time-frequency spectrograms; in contrast, we use a minimal five-channel set and a direct signal-to-grayscale encoding, and we combine residual feature extraction with SE attention in a single lightweight model.

\subsection{Sleep-Deprivation Detection}
Deep learning has been applied to detect sleep deprivation from EEG, but most studies address binary within-subject classification (deprived vs. rested) using full-channel montages and report accuracies in the 90--95\% range~\cite{alsolai2024sleep}. Deep learning has likewise become the dominant paradigm for automatic sleep staging, where convolutional and sequence models achieve expert-level scoring from single-channel EEG and generalize across large, heterogeneous cohorts~\cite{supratak2017deepsleepnet, perslev2021usleep}. Lightweight and deployable models are increasingly emphasized for portable monitoring~\cite{phan2024sleeplight}, while cross-subject generalization remains limited by inter-subject variability~\cite{wu2024transfer}. Our work differs by jointly classifying the sleep condition (NS vs. SD) and the eye state (EO vs. EC) as a single four-class problem under a reduced electrode configuration, a setting not addressed by the binary formulations above.

\section{Methodology}
\label{sec:methodology}
\subsection{Dataset Description}
\label{subsec:dataset}
The EEG dataset used in this study was collected from 71 subjects and stored in the Brain Imaging Data Structure (BIDS) format as \texttt{.set} files. The recordings were acquired using a 61-channel EEG system at a sampling rate of 500~Hz. Each subject participated in two experimental sessions: Session~1 (ses-1) corresponding to normal sleep (NS) condition, and Session~2 (ses-2) corresponding to the sleep deprivation (SD) condition. Within each session, subjects performed resting-state tasks with alternating eyes-open (EO) and eyes-closed (EC) periods, yielding four distinct classes: ses-1\_EyesOpen (EO-NS), ses-1\_EyesClosed (EC-NS), ses-2\_EyesOpen (EO-SD), and ses-2\_EyesClosed (EC-SD).
Due to variability in data quality across subjects, a subject selection procedure was applied. Specifically, 35 subjects (sub-2 through sub-38, non-consecutive) were retained based on signal quality criteria, while the remaining subjects were excluded due to missing data or excessive artifacts.

\subsection{EEG Preprocessing and Channel Selection}
\label{subsec:preprocessing}
\begin{figure*}[!t]
    \centering
    \includegraphics[width=\textwidth]{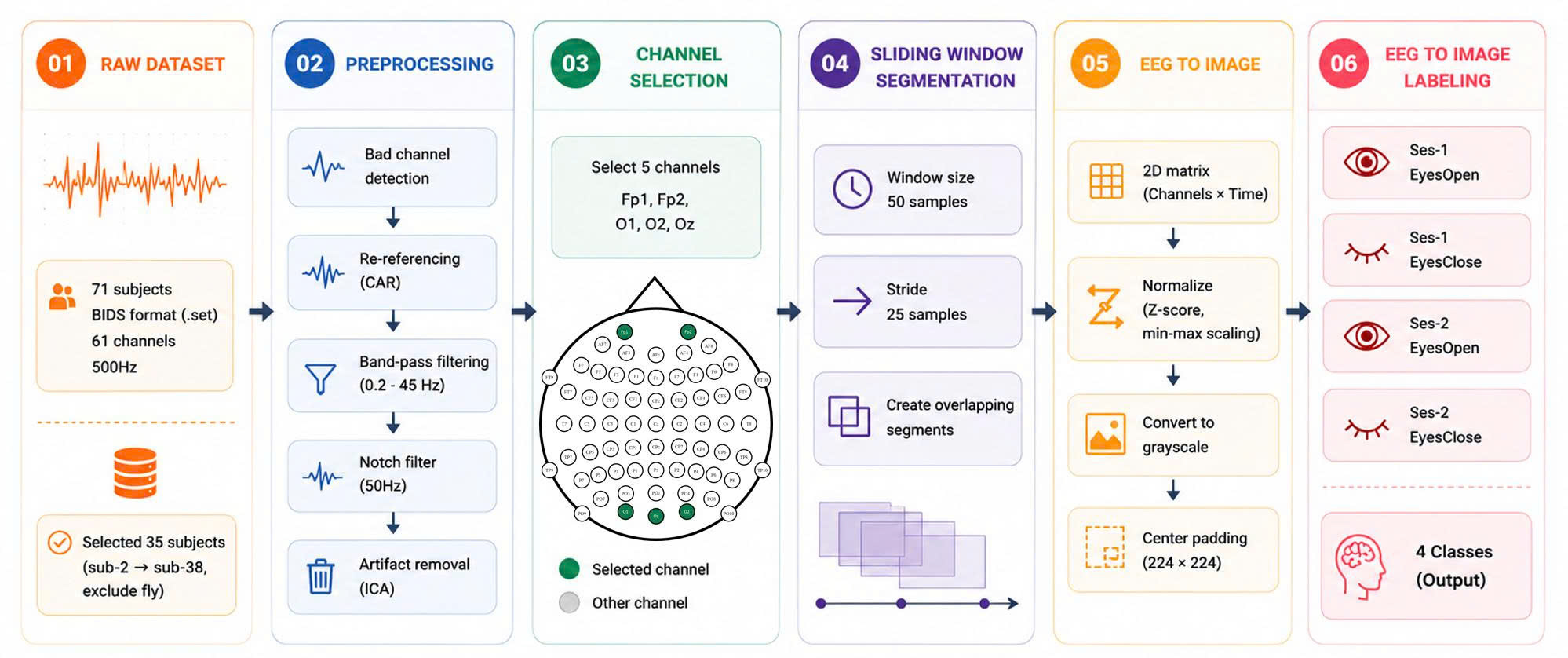}
    \caption{Overview of the proposed EEG data processing pipeline. The pipeline consists of six stages: (1)~raw dataset acquisition from 71 subjects with 61-channel EEG at 500~Hz in BIDS format; (2)~preprocessing including bad channel detection, re-referencing, band-pass filtering (0.2--45~Hz), notch filtering (50~Hz), and ICA-based artifact removal; (3)~channel selection reducing to 5 channels (Fp1, Fp2, O1, Oz, O2); (4)~sliding window segmentation with window size of 50 samples and stride of 25 samples; (5)~EEG-to-image conversion via Z-score normalization, min-max scaling, grayscale conversion, and center padding to $224\times224$; (6)~output labeling into four classes.}
    \label{fig:methodology}
\end{figure*}

The overall data processing pipeline is illustrated in Fig.~\ref{fig:methodology}. A reference preprocessing pipeline from the original dataset study was considered, comprising band-pass filtering (0.2--45~Hz), notch filtering (50~Hz), bad channel detection, re-referencing using the common average reference (CAR), and artifact removal using Independent Component Analysis (ICA).
Following preprocessing, channel dimensionality was reduced by selecting five representative EEG channels: Fp1, Fp2, O1, Oz, and O2. This selection targets the frontal (Fp1, Fp2) and occipital (O1, Oz, O2) regions, which are well-established as the primary cortical areas modulated by eye state and sleep deprivation~\cite{barry2007eeg, hoedlmoser2024alpha}. The frontal channels capture prefrontal activity associated with executive function and attention, while the occipital channels capture the dominant alpha rhythms that characterize the eyes-closed state and their suppression during eyes-open conditions. This strategic reduction from 61 to 5 channels substantially decreases computational complexity while preserving the most discriminative neural features.

\subsection{Sliding Window Segmentation}
\label{subsec:segmentation}
The continuous EEG signals were segmented into fixed-length epochs using a sliding window approach. A window size of 50 samples (corresponding to 100~ms at 500~Hz) with a stride of 25 samples (50\% overlap) was applied to each channel independently. Each resulting segment had dimensions of $50 \times 5$ (time samples $\times$ channels), capturing short-duration temporal dynamics across the selected channels.

Because a 50\% overlap causes two consecutive windows to share half of their samples, performing the train--validation--test split after window segmentation would allow highly correlated windows to appear in different data partitions, leading to information leakage and an overestimation of classification performance. To eliminate this issue, the continuous EEG recording of each subject was first partitioned chronologically into mutually exclusive training, validation, and test intervals before any preprocessing that generated overlapping samples. The overlapping sliding-window procedure described in Section~\ref{subsec:segmentation} was subsequently applied independently within each partition. Therefore, no overlapping window crosses the train--validation or train--test boundaries, and no sample is shared between different data partitions. This leakage-safe design ensures that the observed performance reflects the model's true ability to generalize to unseen temporal data rather than benefiting from duplicated information introduced by overlapping windows.

\subsection{EEG-to-Image Conversion}
\label{subsec:eeg_to_image}
Each $50 \times 5$ segment was transformed into a two-dimensional grayscale image through a multi-step normalization and encoding process:
\begin{enumerate}
    \item \textbf{Z-score normalization}: Per-channel Z-score normalization was applied to remove mean and variance biases, ensuring standardized signal amplitudes across channels and subjects.
    \item \textbf{Min-max scaling}: The normalized values were rescaled to the $[0, 255]$ range using min-max scaling to conform to the grayscale image intensity range.
    \item \textbf{Center padding}: Each $50 \times 5$ matrix was embedded at the center of a $224 \times 224$ canvas with zero-padding, producing input images compatible with standard CNN architectures. This padding strategy preserves the original signal structure at the center while providing a uniform input size.
\end{enumerate}
The resulting dataset comprised grayscale PNG images categorized into four classes corresponding to the experimental conditions. This image-based representation enables the application of powerful 2D CNN architectures originally designed for computer vision tasks to EEG classification.

\subsection{TinyCNNDeep Architecture}
\label{subsec:architecture}
\begin{figure*}[!t]
    \centering
    \includegraphics[width=\textwidth]{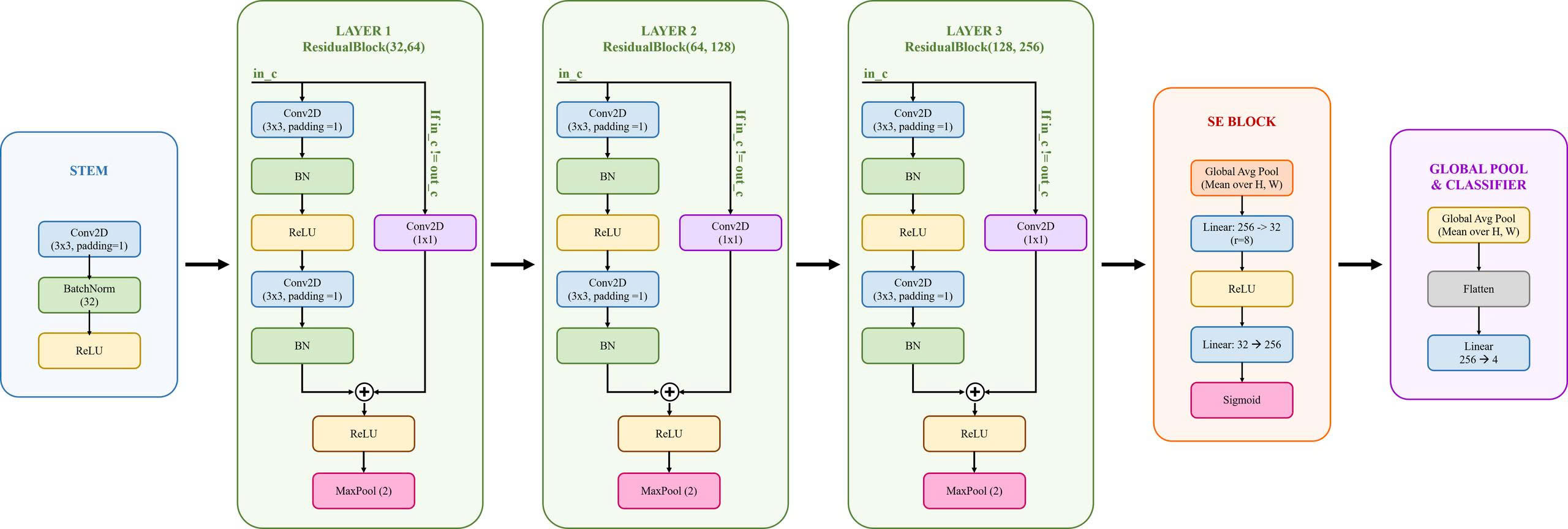}
    \caption{Detailed architecture of the proposed TinyCNNDeep model. The model receives $224\times224\times1$ grayscale EEG images as input and processes them through a Stem Block (Conv2D $3\times3$, 32 filters, BN, ReLU), three Residual Blocks with progressive channel expansion (64, 128, 256) and skip connections, an SE Attention Block for channel-wise recalibration (reduction ratio $r=8$), and a Classification Head (Adaptive Average Pooling, Flatten, FC $256\rightarrow4$) producing four-class output probabilities.}
    \label{fig:architecture}
\end{figure*}

The proposed TinyCNNDeep architecture is illustrated in Fig.~\ref{fig:architecture}. It is designed as a lightweight yet effective model for EEG image classification, comprising four main components: a stem block, three residual blocks, an SE attention module, and a classification head.
\subsubsection{Stem Block}
The stem block serves as the initial feature extractor, consisting of a $3\times3$ convolutional layer with 32 filters (padding=1), followed by Batch Normalization (BN)~\cite{ioffe2015batch} and ReLU activation. This block extracts low-level features from the input grayscale images, producing feature maps of size $224\times224\times32$.
\subsubsection{Residual Blocks}
Three residual blocks are stacked sequentially, each implementing identity shortcut connections to facilitate gradient flow and enable deeper feature learning without degradation~\cite{he2016deep}:
\begin{itemize}
    \item \textbf{Residual Block~1} ($32 \rightarrow 64$): Two $3\times3$ convolutional layers with 64 filters, a $1\times1$ convolution for channel adjustment in the skip connection, BN, and ReLU activation. Output: $224\times224\times64$.
    \item \textbf{Residual Block~2} ($64 \rightarrow 128$): Same structure with 128 filters, followed by MaxPooling with factor 2. Output: $112\times112\times128$.
    \item \textbf{Residual Block~3} ($128 \rightarrow 256$): Same structure with 256 filters, followed by MaxPooling with factor 2. Output: $56\times56\times256$.
\end{itemize}
The progressive channel expansion (64$\rightarrow$128$\rightarrow$256) coupled with spatial downsampling enables hierarchical feature abstraction, capturing increasingly complex patterns in the EEG image representation.
\subsubsection{Squeeze-and-Excitation (SE) Attention Block}
Following the residual blocks, an SE module~\cite{hu2018squeeze} is integrated to enhance channel-wise feature selectivity. The SE block operates as follows:
\begin{enumerate}
    \item \textbf{Squeeze}: Global Average Pooling (GAP) compresses spatial dimensions, producing a channel descriptor vector of size $1\times1\times256$.
    \item \textbf{Excitation}: Two fully connected (FC) layers with a reduction ratio $r=8$ (256$\rightarrow$32$\rightarrow$256) learn inter-channel dependencies. ReLU activation is applied after the first FC layer, and Sigmoid activation after the second.
    \item \textbf{Recalibration}: The resulting channel weights are multiplied element-wise with the original feature maps, selectively emphasizing informative channels and suppressing less relevant ones.
\end{enumerate}
\subsubsection{Classification Head}
The classification head comprises Adaptive Average Pooling (reducing spatial dimensions to $1\times1$), a Flatten operation, and a single FC layer mapping from 256 dimensions to 4 output classes. The output represents the probability distribution over the four EEG states: EO-NS, EC-NS, EO-SD, and EC-SD.

\subsection{Training Configuration}
\label{subsec:training}
The model was trained and evaluated under a subject-wise protocol: a separate model was fit for each subject, and the reported accuracy is the mean over the 35 subjects. This protocol measures intra-subject classification performance and characterizes inter-subject variability, but it does not assess cross-subject generalization (Section~\ref{sec:discussion}).

\subsubsection{Leakage-Safe Data Partitioning}
For each subject, the continuous EEG recording was first divided chronologically into a training set (80\%, consisting of the earlier temporal portion) and a held-out test set (20\%, consisting of the later temporal portion). From the training set, 10\% of the samples were further reserved as a validation set for hyperparameter tuning and early stopping. As described in Section~\ref{subsec:segmentation}, the overlapping sliding-window segmentation was applied independently within the training, validation, and test sets after temporal partitioning. Consequently, overlapping windows never crossed the train--validation or train--test boundaries. This leakage-safe protocol effectively prevents information leakage caused by shared samples between overlapping windows and ensures that model evaluation reflects genuine generalization to unseen temporal data rather than artificially inflated performance due to window overlap.

\subsubsection{Optimization}

The proposed network was trained by minimizing the multi-class cross-entropy loss using the Adam optimizer~\cite{kingma2015adam}. The initial learning rate was set to $1\times10^{-4}$ with $\beta_1=0.9$, $\beta_2=0.999$, and a weight decay of $1\times10^{-5}$. Training was performed with a batch size of 32 for a maximum of 100 epochs. Early stopping based on the validation loss with a patience of 10 epochs was employed to mitigate overfitting, and the model checkpoint achieving the lowest validation loss was selected for final evaluation on the held-out test set.

To address potential class imbalance, weighted cross-entropy loss was adopted, where class weights were computed inversely proportional to the class frequencies in the training set. All experiments were conducted using a fixed random seed of 42 to ensure reproducibility, and the reported results correspond to the average performance over five independent runs.

The implementation was developed in PyTorch 2.12 and executed on a workstation equipped with an NVIDIA GeForce RTX 3090 GPU.

\subsection{Baseline Models}
\label{subsec:baselines}
To comprehensively evaluate the proposed approach, TinyCNNDeep was benchmarked against two categories of baseline models across three feature domains.
\paragraph{Deep Learning Baselines}
\begin{itemize}
    \item \textbf{EEGNet}~\cite{lawhern2018eegnet}: A compact CNN using depthwise and separable convolutions.
    \item \textbf{ShallowConvNet}~\cite{schirrmeister2017deep}: Designed to mimic filter bank common spatial patterns.
    \item \textbf{DeepConvNet}~\cite{schirrmeister2017deep}: A deeper architecture for general EEG decoding.
\end{itemize}
\paragraph{Classical Machine Learning Baselines}
\begin{itemize}
    \item Support Vector Machine (SVM)
    \item Random Forest
    \item k-Nearest Neighbors (KNN)
    \item Gradient Boosting
    \item Logistic Regression
\end{itemize}
Each baseline was evaluated on three feature representations: time-domain features, frequency-domain features, and combined time+frequency domain features.

\section{Results}
\label{sec:results}
\subsection{Subject-Wise Classification Performance}
\label{subsec:subject_results}
Table~\ref{tab:subject_results} presents the subject-wise mean accuracy of the proposed TinyCNNDeep model across all 35 subjects. The model achieved an overall mean accuracy of \textbf{83.69\%} across all subjects, with substantial variation reflecting inter-subject differences in EEG signal characteristics.
Several subjects achieved near-perfect classification accuracy, including sub-19 (99.97\%), sub-30 (99.95\%), sub-36 (99.95\%), sub-32 (99.79\%), sub-14 (99.76\%), sub-33 (99.73\%), sub-35 (99.65\%), sub-16 (99.60\%), and sub-6 (99.44\%). These results demonstrate the model's capacity to capture highly discriminative EEG patterns for subjects with well-separated class distributions.
Conversely, some subjects exhibited lower accuracy, with sub-22 (50.48\%), sub-17 (56.28\%), sub-21 (58.72\%), and sub-12 (60.35\%) representing the most challenging cases. These lower-performing subjects likely reflect higher intra-subject noise, atypical EEG patterns, or less distinct differences between the four experimental conditions.
\begin{table}[!t]
\centering
\caption{Subject-Wise Classification Accuracy (\%) of TinyCNNDeep.}
\label{tab:subject_results}
\renewcommand{\arraystretch}{1.1}
\begin{tabular}{lc|lc}
\toprule
\textbf{Subject} & \textbf{Accuracy (\%)} & \textbf{Subject} & \textbf{Accuracy (\%)} \\
\midrule
sub-2  & 96.09 & sub-22 & 50.48 \\
sub-4  & 86.99 & sub-23 & 77.15 \\
sub-5  & 71.22 & sub-24 & 68.94 \\
sub-6  & 99.44 & sub-25 & 81.38 \\
sub-7  & 97.02 & sub-26 & 79.71 \\
sub-8  & 66.30 & sub-27 & 85.45 \\
sub-9  & 89.10 & sub-29 & 91.89 \\
sub-10 & 64.63 & sub-30 & 99.95 \\
sub-11 & 70.64 & sub-31 & 99.04 \\
sub-12 & 60.35 & sub-32 & 99.79 \\
sub-13 & 79.47 & sub-33 & 99.73 \\
sub-14 & 99.76 & sub-34 & 93.72 \\
sub-15 & 79.68 & sub-35 & 99.65 \\
sub-16 & 99.60 & sub-36 & 99.95 \\
sub-17 & 56.28 & sub-37 & 92.13 \\
sub-18 & 70.90 & sub-38 & 78.38 \\
sub-19 & 99.97 &        &       \\
sub-20 & 85.61 &        &       \\
sub-21 & 58.72 &        &       \\
\midrule
\multicolumn{2}{c|}{\textbf{Overall Mean}} & \multicolumn{2}{c}{\textbf{83.69}} \\
\bottomrule
\end{tabular}
\end{table}

\subsection{Benchmark Comparison}
\label{subsec:benchmark}
Table~\ref{tab:benchmark} presents the comprehensive benchmark comparison of TinyCNNDeep against all baseline models across three feature domains. The results show that the proposed approach attains substantially higher accuracy than all baselines.
Among the deep learning baselines, all three models (EEGNet, ShallowConvNet, DeepConvNet) performed near chance level ($\sim$25\% for a 4-class problem) across all feature domains. In the time domain, EEGNet achieved 28.40\%, ShallowConvNet 25.43\%, and DeepConvNet 25.89\%. Similar poor performance was observed in the frequency domain (23.14\%--27.31\%) and the combined time+frequency domain (25.37\%--27.09\%).
The classical machine learning baselines performed moderately better but still substantially below TinyCNNDeep. Among all baselines and feature domains, Random Forest with combined time+frequency features achieved the highest accuracy at 47.66\%, followed by Gradient Boosting at 46.17\% and Logistic Regression at 44.40\% in the same feature domain. SVM achieved 44.29\%, while KNN reached 40.51\%.
TinyCNNDeep's mean accuracy of \textbf{83.69\%} thus exceeds the best-performing baseline (Random Forest, time+frequency) by a margin of \textbf{36.03 percentage points}, representing a \textbf{75.6\% relative improvement}.
\begin{table*}[!t]
\centering
\caption{Benchmark Comparison of Mean Classification Accuracy (\%) Across Different Models and Feature Domains. The Best Result in Each Domain is Shown in \textbf{Bold}.}
\label{tab:benchmark}
\renewcommand{\arraystretch}{1.15}
\begin{tabular}{llccc}
\toprule
\textbf{Category} & \textbf{Model} & \textbf{Time Domain} & \textbf{Frequency Domain} & \textbf{Time + Frequency} \\
\midrule
\multirow{3}{*}{Deep Learning}
& EEGNet         & 28.40 & 23.14 & 27.09 \\
& ShallowConvNet & 25.43 & 24.34 & 25.71 \\
& DeepConvNet    & 25.89 & 27.31 & 25.37 \\
\midrule
\multirow{5}{*}{\shortstack[l]{Classical\\Machine Learning}}
& SVM                  & 42.34 & 41.60 & 44.29 \\
& Random Forest        & 45.03 & 44.11 & \textbf{47.66} \\
& KNN                  & 35.89 & 40.23 & 40.51 \\
& Gradient Boosting    & 43.71 & 40.29 & 46.17 \\
& Logistic Regression  & 42.40 & 41.20 & 44.40 \\
\midrule
\textbf{Proposed} & \textbf{TinyCNNDeep (Image-based)} & \multicolumn{3}{c}{\textbf{83.69}} \\
\bottomrule
\end{tabular}
\end{table*}

\subsection{Confusion Matrix Analysis}
\label{subsec:confusion}
\begin{figure}[!t]
    \centering
    \includegraphics[width=0.85\columnwidth]{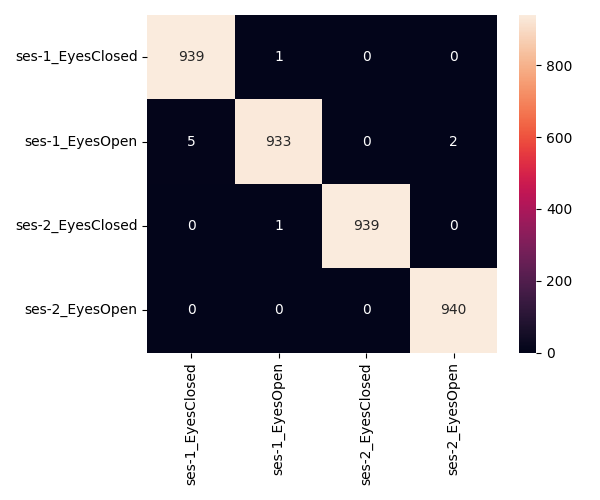}
    \caption{Confusion matrix for sub-14 (accuracy: 99.76\%), representing a high-performing subject with near-perfect discrimination across all four classes.}
    \label{fig:confusion_good}
\end{figure}

\begin{figure}[!t]
    \centering
    \includegraphics[width=0.85\columnwidth]{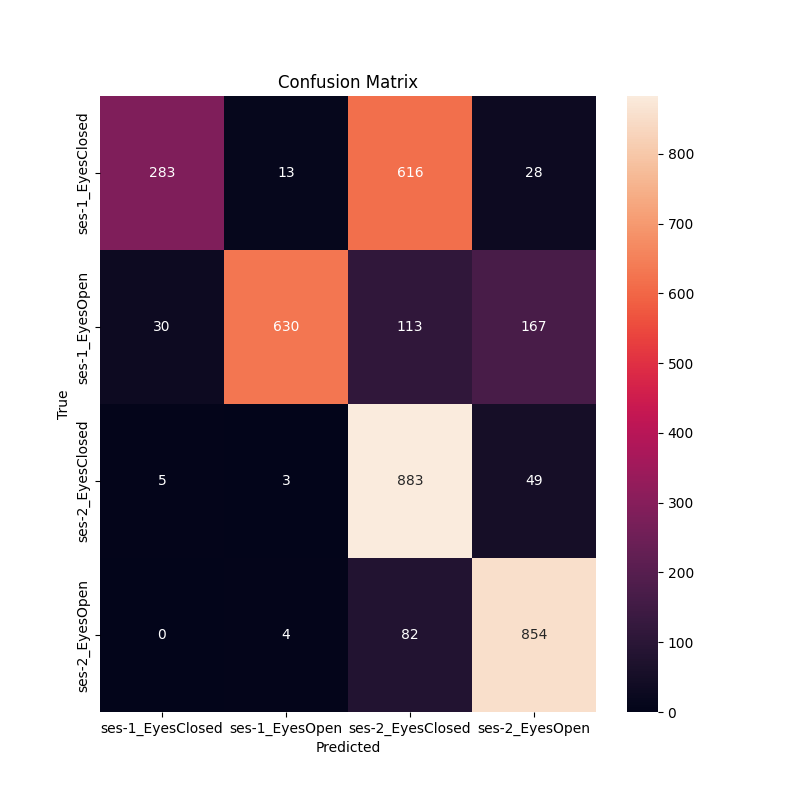}
    \caption{Confusion matrix for sub-5 (accuracy: 71.22\%), illustrating inter-class confusion patterns, particularly between eyes-closed states across normal sleep and sleep deprivation conditions.}
    \label{fig:confusion_challenging}
\end{figure}

Fig.~\ref{fig:confusion_good} and Fig.~\ref{fig:confusion_challenging} present representative confusion matrices for a high-performing subject (sub-14) and a more challenging subject (sub-5), respectively. For sub-14, the confusion matrix shows near-perfect classification across all four classes, with only 9 misclassified samples out of 3,760 total, indicating clear discrimination between both eye states and sleep conditions. In contrast, sub-5 shows notable confusion patterns, particularly between ses-1\_EyesClosed and ses-2\_EyesClosed (616 misclassifications), suggesting that for some subjects, the EEG patterns associated with the eyes-closed state are similar regardless of sleep condition.

\section{Discussion}
\label{sec:discussion}

\subsection{Image-Based EEG Representation}
The most striking finding of this study is the dramatic performance gap between TinyCNNDeep (83.69\%) and all baseline models ($\leq$47.66\%). This disparity can be attributed to several factors. The image-based representation preserves the inherent two-dimensional structure of multi-channel EEG data (channels $\times$ time), enabling 2D convolutions to jointly learn spatial (inter-channel) and temporal (intra-channel) features, an advantage lost when signals are flattened into one-dimensional feature vectors for classical machine learning~\cite{bashivan2016learning}. Furthermore, the normalization and image encoding pipeline (Z-score followed by min-max scaling to grayscale) standardizes the data across subjects and channels, mitigating amplitude variability that often degrades conventional classifiers.

The near-chance performance of EEGNet, ShallowConvNet, and DeepConvNet (23\%--28\% for a 4-class problem) is particularly noteworthy. These architectures, originally designed for raw EEG time-series classification with temporal and spatial filter banks, appear ill-suited for the four-class classification task under a reduced five-channel configuration. Recent benchmarking studies have similarly reported that these models exhibit inconsistent performance across different datasets and evaluation paradigms, with effectiveness being highly dependent on training data volume and the specific EEG task~\cite{hernandez2024eeg}. Our findings reinforce that established EEG-specific architectures do not universally generalize across all classification scenarios, and that task-specific architectural design, such as the image-based approach proposed here, can produce substantially higher accuracy.

Classical machine learning methods (SVM, Random Forest, Gradient Boosting, Logistic Regression, KNN) performed moderately better than the deep learning baselines (35\%--48\%), consistent with recent observations that for certain EEG applications with limited channels or noisy data, traditional feature-engineered approaches can outperform standard deep learning architectures~\cite{wu2024transfer}. Nevertheless, their accuracy remained far below TinyCNNDeep, indicating the limitations of manually designed features for capturing the complex, non-linear patterns that distinguish four closely related EEG states.

\subsection{Role of Residual Learning and SE Attention}
The architectural design choices of TinyCNNDeep contribute meaningfully to its strong performance. The residual connections, inspired by ResNet~\cite{he2016deep}, address the degradation problem in deep networks by enabling identity mappings that facilitate gradient flow during training. This is particularly important for EEG classification tasks where the informative signal content occupies a small fraction of the input image (a $50\times5$ signal centered within a $224\times224$ canvas), requiring the network to learn effectively from sparse, localized features.

The SE attention module provides an additional performance advantage through adaptive channel-wise feature recalibration. In the context of EEG classification, different convolutional feature channels may encode distinct neurophysiological signatures, such as alpha suppression patterns, frontal theta activity, or sleep-deprivation-induced spectral shifts. The SE block dynamically weights these features based on their discriminative relevance for each input, implementing a learned attention mechanism over the feature space~\cite{hu2018squeeze, li2024se_eeg}. This is consistent with recent findings that attention mechanisms substantially improve EEG classification by focusing on clinically relevant signal components~\cite{zhang2025se_bci}. The lightweight nature of the SE module (reduction ratio $r=8$, adding minimal parameters) makes it especially attractive for resource-constrained applications, aligning with the broader trend toward efficient, deployable BCI models~\cite{phan2024sleeplight}.

\subsection{Inter-Subject Variability Analysis}
The substantial range in subject-wise accuracy (50.48\%--99.97\%) highlights the well-documented challenge of inter-subject variability in EEG research~\cite{wu2024transfer, tran2026variability}. High-performing subjects (e.g., sub-19: 99.97\%, sub-14: 99.76\%, sub-30: 99.95\%) likely exhibit clear, consistent neurophysiological differences between the four conditions, such as pronounced alpha rhythm modulation between EO and EC states and distinct spectral signatures between NS and SD sessions.

In contrast, lower-performing subjects (e.g., sub-22: 50.48\%, sub-17: 56.28\%) may exhibit attenuated or irregular EEG patterns, potentially reflecting individual differences in cortical excitability, habitual sleep patterns, or varying degrees of physiological adaptation to sleep deprivation~\cite{hoedlmoser2024alpha}. The confusion matrix analysis (Fig.~\ref{fig:confusion_challenging}) shows that misclassifications predominantly occur between same-eye-state classes across different sleep conditions (e.g., EC-NS vs. EC-SD), rather than between eye states within the same condition. This suggests that for some subjects, sleep deprivation induces relatively subtle EEG changes that are insufficient to reliably distinguish NS from SD within a given eye state, while the EO/EC distinction remains more robust, consistent with the strong alpha modulation associated with eye closure~\cite{barry2007eeg}.

\subsection{Neurophysiological Interpretation}
The selection of frontal (Fp1, Fp2) and occipital (O1, Oz, O2) channels is grounded in established neurophysiology. Occipital regions are the primary generators of alpha rhythms that characterize the eyes-closed resting state~\cite{barry2007eeg}, and recent research has shown that sleep deprivation induces significant changes in aperiodic activity and alpha bandwidth specifically in occipital regions~\cite{cassim2024aperiodic}. Frontal channels capture prefrontal activity related to executive function and vigilance, regions known to be particularly vulnerable to sleep deprivation effects including increased delta and theta power~\cite{killgore2010effects, hoedlmoser2024alpha}.

The model's ability to achieve high accuracy with only five channels demonstrates that the critical information for distinguishing between the four EEG states is concentrated in these physiologically relevant regions. This finding has practical implications for the design of simplified, portable EEG systems targeting sleep monitoring applications, where minimizing the number of electrodes is essential for user comfort and practical deployment~\cite{phan2024sleeplight}.

\subsection{Comparison with Existing Literature}
Compared to recent studies on EEG-based sleep deprivation detection, which typically report 90--95\% accuracy for binary (sleep-deprived vs. rested) within-subject classification using full-channel configurations~\cite{alsolai2024sleep}, our four-class accuracy of 83.69\% with only five channels represents a competitive result given the substantially more challenging task formulation. While binary sleep state classification is inherently simpler, our model simultaneously discriminates both the sleep condition axis (NS vs. SD) and the eye state axis (EO vs. EC), providing richer and more clinically useful information.

The image-based approach adopted here aligns with the growing trend of transforming EEG data into visual representations for CNN processing~\cite{bashivan2016learning, wang2024eeg_image}, but differs from prior work in several key aspects: (1) we use a minimal channel set rather than full-scalp topographic maps; (2) we employ direct signal-to-grayscale conversion rather than time-frequency spectrograms; and (3) we target a four-class problem that jointly captures sleep condition and eye state. These design choices prioritize computational efficiency and practical applicability while maintaining strong classification performance.

\subsection{Limitations and Future Directions}
Several limitations of this study warrant consideration. First, the subject-wise evaluation protocol, while characterizing inter-subject variability, does not assess cross-subject generalization. Future work should investigate transfer learning and domain adaptation techniques to improve performance on unseen subjects~\cite{wu2024transfer}. Second, the center-padding strategy, while effective, results in a large proportion of zero-valued pixels in the input images. Alternative image encoding strategies, such as tiling, interpolation, or time-frequency representations (spectrograms, scalograms), may yield more information-dense inputs and potentially improve performance.

Third, the current study does not explore model interpretability techniques such as Grad-CAM or attention visualization, which could provide valuable insights into which spatial and temporal features drive classification decisions. Fourth, the sliding window parameters (window size = 50, stride = 25) were fixed; systematic optimization of these hyperparameters through ablation studies may further enhance performance.

Future research directions include: (1) incorporating cross-subject validation and transfer learning to improve generalizability; (2) exploring alternative EEG-to-image encoding methods such as continuous wavelet transforms or short-time Fourier transform spectrograms; (3) integrating explainability tools to enhance clinical interpretability; (4) extending the approach to additional sleep-related classification tasks such as sleep stage scoring~\cite{supratak2017deepsleepnet, perslev2021usleep}; and (5) deploying TinyCNNDeep on edge devices to validate real-time performance for wearable sleep monitoring applications.

\section{Conclusion}
\label{sec:conclusion}
This study proposed TinyCNNDeep, a lightweight convolutional neural network integrating residual learning and Squeeze-and-Excitation attention, for the four-class classification of EEG signals across eyes-open and eyes-closed states under normal sleep and sleep deprivation conditions. By converting five-channel EEG segments (Fp1, Fp2, O1, Oz, O2) into $224\times224$ grayscale images through Z-score normalization, min-max scaling, and center padding, the proposed model achieved a mean accuracy of 83.69\% across 35 subjects, with 10 subjects exceeding 99\%, outperforming all baselines by at least 36 percentage points over the best competitor (Random Forest at 47.66\%), while established deep learning models (EEGNet, ShallowConvNet, DeepConvNet) performed near chance level. These results demonstrate that the synergy of image-based EEG representation, residual feature extraction, and channel-wise attention constitutes an effective and computationally efficient framework for multi-class sleep-related EEG classification, even with a minimal electrode configuration suitable for portable monitoring systems. Future work will address cross-subject generalization via transfer learning, explore alternative signal-to-image encodings such as time-frequency spectrograms, integrate model interpretability techniques for clinical validation, and target edge deployment for real-time wearable sleep monitoring.


\balance
\bibliographystyle{IEEEtran}
\bibliography{tinycnndeep}

\end{document}